\documentstyle[prl,aps,preprint,epsfig,floats]{revtex}   

\begin{document}

\draft

\def\d{{\rm d}}
\def\e{{\rm e}}
\def\eref#1{(\protect\ref{#1})}
\def\fref#1{\protect\ref{#1}}
\def\etal{{\it{}et~al.}}
\def\set#1{\lbrace#1\rbrace}
\def\av#1{\langle#1\rangle}

\def\capt#1{\refstepcounter{figure}\bigskip\hbox to \textwidth{%
       \hfil\vbox{\hsize=5in\renewcommand{\baselinestretch}{1}\small
       {\sc Figure \thefigure}\quad#1}\hfil}\bigskip}


\renewcommand{\baselinestretch}{1.1}

\title{The Physical Limits of Communication\\
{\em --- or ---}\\
Why any sufficiently advanced technology is\\
indistinguishable from noise}
\author{Michael Lachmann, M. E. J. Newman, and Cristopher Moore}
\address{Santa Fe Institute, 1399 Hyde Park Road, Santa Fe, NM 87501.
U.S.A.}
\maketitle
\vspace{1cm}
\begin{abstract}
  It has been well-known since the pioneering work of Claude Shannon in the
  1940s that a message transmitted with optimal efficiency over a channel
  of limited bandwidth is indistinguishable from random noise to a receiver
  who is unfamiliar with the language in which the message is written.  In
  this letter we demonstrate an equivalent result about electromagnetic
  transmissions.  We show that when electromagnetic radiation is used as
  the transmission medium, the most information-efficient format for a
  given message is indistinguishable from black-body radiation to a
  receiver who is unfamiliar with that format.  The characteristic
  temperature of the radiation is set by the amount of energy used to make
  the transmission.  If information is not encoded in the direction of the
  radiation, but only its timing, energy or polarization, then the most
  efficient format has the form of a one-dimensional black-body spectrum
  which is easily distinguished from the three-dimensional case.
\end{abstract}

\newpage


\renewcommand{\baselinestretch}{1.5}

Shannon's information theory~\cite{Shannon49,CT91} considers the set
$\set{x_i}$ of all possible messages $x_i$ which can be transmitted across
a ``channel'' linking a sender to a receiver.  In the simplest case the
channel is considered to be free of noise, so that each message is received
just as it was sent.  Shannon demonstrated that the only consistent
definition of the amount of information carried by such a channel is
\begin{equation}
S = -\sum_i p_i \log p_i,
\label{shannon}
\end{equation}
where $p_i$ is the probability of transmission of the message $x_i$.  (In
evaluating this expression we will take natural logarithms, as is common in
statistical mechanics.  In information theory logarithms base~2 are the
norm, but the difference is only one of a constant factor in $S$.
Conventionally in statistical mechanics there is also a leading constant
$k$ in the definition of $S$, known as the Boltzmann constant.  Here we
have set $k=1$, with the result that temperatures are measured in units of
energy.)

If there are no other constraints, then $S$ is maximized when all the $p_i$
are equal, i.e.,~when all messages are transmitted with equal probability.
If we transmit a constant stream of information in this way by sending many
messages one after another, then the resulting flow of data will appear
completely random to anyone who does not know the language that it is
written in.

We now consider the corresponding problem for a message sent using
electromagnetic radiation.  In outline our argument is as follows.  We
assume that the sender of a radio message has a certain amount of energy
available and we ask what is the maximum amount of information that can be
sent with that energy.  Generically, this is the problem of maximizing the
Shannon information, Eq.~\eref{shannon}, for an ensemble of bosons (photons
in this case) subject to the constraint of fixed average energy.  The
solution is familiar from statistical physics, since the formula for the
Shannon information is identical to that for the thermodynamic entropy of
an ensemble.  The maximization of entropy for an ensemble of bosons gives
rise to the Bose--Einstein statistics and, in the case of electromagnetic
waves, to black-body radiation.  By an exact analogy we now show that the
most information-rich electromagnetic transmission has a spectrum
indistinguishable from black-body radiation.

In order to apply Shannon's theory to electromagnetic radiation we must
pose the problem in a form resembling the transmission of information over
a channel.  To do that, consider the following thought experiment.  Suppose
we have a closed container or ``cavity'' with perfectly reflecting walls,
which, for reasons which will shortly become clear, we take to be a long
tube of constant cross-sectional area $A_t$ and length $\ell$.  Suppose
also that we have the technical wherewithal to set up any electromagnetic
``microstate'' within this cavity.  (The microstates need not be single
modes of the cavity; any set of states will work.)  Since the walls are
perfectly reflecting, this microstate will remain in place indefinitely.
We can use the cavity as a communication device.  Each possible message
$x_i$ that we might wish to send is agreed to correspond to one microstate,
and we simply pass the entire cavity to the receiver of the message.  The
information transferred by the electromagnetic radiation stored within is
then given by Shannon's formula with $p_i$ being the probability that we
put the cavity in microstate $i$.  (This communication is not the same as a
sending a radio message but does have the same information content, as we
will demonstrate shortly.)

Now suppose that, as in the case of the simple channel, we wish to transmit
a steady stream of information by exchanging a series of such cavities.  We
assume that the power available to make the transmission is
limited---i.e.,~there is a fixed average energy ``budget'' $\av{E}$ which
we can spend per message sent---and we ask what is the greatest amount of
information that can be transmitted per message.  This is given by
maximizing Eq.~\eref{shannon} subject to the constraint of fixed average
energy, but allowing any number of photons to be present in the cavity.
The solution of this maximization problem is
well-known~\cite{Jaynes83,Grandy87} and gives the grand canonical ensemble
in which
\begin{equation}
p_i = {\exp(-\beta E_i - \gamma N_i)\over Z},
\end{equation}
where $E_i$ is the energy in microstate $i$, $N_i$ is the number of
photons, $Z$ is the grand canonical partition function, and $\beta$ and
$\gamma$ are Lagrange multipliers, usually referred to as the (inverse)
temperature and chemical potential respectively.  If we now denote the
microstates by the numbers of photons $\set{n_k}$ in each (non-interacting)
single-particle state $k$ of the cavity, then it is straightforward to show
that the average of any $n_k$ over many messages follows the normal
Bose--Einstein distribution
\begin{equation}
\av{n_k} = {1\over\e^{\beta\varepsilon_k}-1},
\label{nstates}
\end{equation}
where $\varepsilon_k$ is the energy of a photon in state $k$ and we have
set $\gamma=0$ since there is no chemical potential for photons in a
vacuum.

\begin{figure}
\begin{center}
\psfig{figure=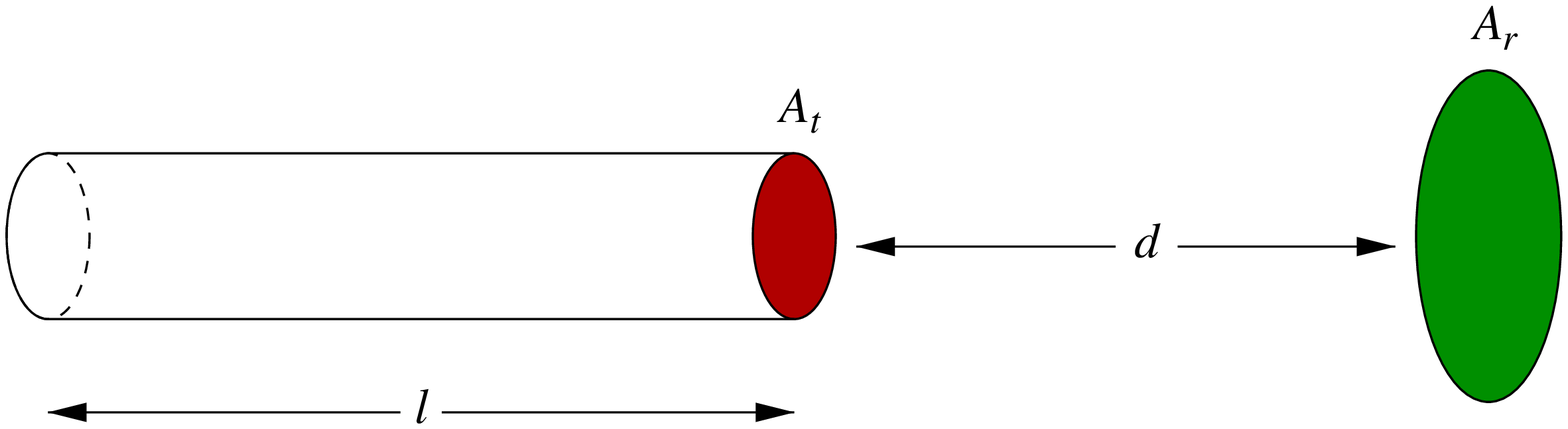,width=5in}
\end{center}
\capt{The setup used in our thought experiment.  A tubular cavity of
  cross-section $A_t$ is used as a transmitter.  The message is received by
  a receiver of area $A_r$ a distance $d$ away.}
\label{setup}
\end{figure}

Suppose now that, instead of handing over the entire cavity, we remove one
of its ends and allow the photons to escape in the form of a message.  One
should envisage the end of the cavity as being the transmitter of the
message, not the cavity itself.  The cross-section $A_t$ of the cavity is
also the area of the transmitter, hence the subscript $t$.  Usually we will
want to direct the message towards a specific receiver.  For concreteness
we will assume that the receiver is a distance $d$ from the end of the
cavity and presents some finite area $A_r$ to the cavity as depicted in
Fig.~\fref{setup}.  Only those photons in the cavity which have their
momentum within the correct interval of solid angle will strike the
receiver, possibly after reflecting one or more times off the walls of the
container.  (We use only the forward interval of solid angle, disallowing
reflections off the rear wall of the cavity.  If we allow such reflections
the transmission will convey twice as much information but last twice as
long, making the rate of information transfer identical.)  Given that our
cavity has volume $V=\ell A_t$, the density of single-particle states
satisfying this criterion, taking into account both polarizations of the
photons, is $\rho(\varepsilon) = 2 \ell A_t A_r \varepsilon^2/d^2 h^3 c^3$,
where $h$ is Planck's constant and $c$ is the speed of light.  The average
energy contained in our message per unit interval of energy is thus
\begin{equation}
I(\varepsilon) = {2 \ell A_t A_r\over d^2 h^3 c^3}\>
{\varepsilon^3\over\e^{\beta\varepsilon}-1}.
\label{intensity}
\end{equation}
A spectrum with this functional form for the intensity $I(\varepsilon)$ is
usually referred to as a black-body spectrum.  A black-body spectrum is
produced by a perfect thermal radiator in equilibrium at temperature $T =
\beta^{-1}$, and, to a good approximation, by most astronomical bodies.

In fact there is a small correction to Eq.~\eref{intensity} because photons
may be delayed slightly if they have to take a path which reflects off the
walls of the cavity.  This correction goes inversely as the cosine of the
angle subtended by the receiver at the transmitter.  Here we have assumed
that the distance between transmitter and receiver is large enough that
this cosine is well approximated by~1.

Our transmission necessarily contains all the information needed to
reconstruct the original microstate of the cavity, and therefore contains
the same amount of information as that microstate.  Furthermore, no
ensemble can exist of messages which originate within a volume less than or
equal to that of our cavity and which convey more information per message
for the same average energy.  If such an ensemble existed, reversing time
would allow us to trap messages drawn from that ensemble within a cavity of
the same size and create an ensemble of microstates which also had greater
information content.  This conclusion is forbidden, since we have already
shown that the black-body distribution maximizes information content.

Now we send a stream of information in the form of many such messages, each
one resulting from one microstate of the cavity.  Because of the tubular
shape we have chosen for our cavity each message will last a time $\ell/c$
and the transmission will have constant average intensity.  The apparent
temperature $T$ of the transmission is fixed by the size of the energy
budget.  Calculating the average energy $\av{E}$ per message from an
integral over Eq.~\eref{intensity} and dividing by $\ell/c$, we find that
for a transmission with energy budget $\epsilon$ per unit time the apparent
temperature is given by
\begin{equation}
T^4 = {15 h^3 c^2\over2\pi^4}\>{d^2\over A_t A_r}\>\epsilon.
\end{equation}
The information transmitted per unit time $\d S/\d t$ can be calculated
using the formula $S =\log Z - \beta\,\partial\log Z/\partial\beta$, noting
that $\partial\log Z/\partial\beta=\av{E}$.  The result is
\begin{equation}
{\d S\over\d t} = {8\pi^4\over45h^3 c^2}\>{A_t A_r\over d^2}\>T^3
  = \Biggl[{512\pi^4\over1215h^3c^2}\>{A_t A_r\over
  d^2}\>\epsilon^3\Biggr]^{1\over4}.
\label{rate}
\end{equation}
This is the greatest possible rate at which information can be transferred
by any electromagnetic transmission for a given average power.  It depends
only on fundamental constants, the areas of the transmitter and receiver,
the distance between them, and the average power or equivalently the
apparent temperature.
For a transmitter and receiver of one square metre each, a metre apart,
with an energy budget of $1\:{\rm Js}^{-1}$, the information rate is
$1.61\times10^{21}$ bits per second.  Note that the information rate
increases slightly slower than linearly with the power $\epsilon$ of the
transmitter, as $\epsilon^{3/4}$.  It also increases with the area of the
transmitter and receiver, so that the best information rate for a given
energy budget is achieved for large antennae with low apparent
temperatures.

In order to reconstruct the microstate of the original cavity and hence
extract the information from a transmission of this kind, we need to have
complete information about the photons arriving at our receiver.  This
means that in addition to energy and polarization information, we need to
know the transverse momentum (or transverse position) of each one.  In
practice, we can at best only detect the transverse components with finite
resolution, which places an upper limit on the energy range over which
Eq.~\eref{intensity} is valid.  The finite size of the receiver places a
lower limit on the range.  Thus, for a given receiver, the most
information-efficient transmission will have a black-body spectrum only
between these limits.

In fact, most receivers are not capable of measuring the transverse
components at all.  Usually we have information about only the energy and
timing of arriving photons, and possibly their polarization.  In this case
Eq.~\eref{nstates} still holds, but Eq.~\eref{intensity} becomes
\begin{equation}
I(\varepsilon) = {2 \ell \over h c}\>
{\varepsilon\over\e^{\beta\varepsilon}-1}.
\label{intensity2}
\end{equation}
This is the form which the spectrum of a black body would take in a
one-dimensional world, and we refer to it as a ``one-dimensional black-body
spectrum''.  The information transmitted per unit time is
\begin{equation}
{\d S\over\d t} = {2\pi^2\over3h}\>T
= {\Biggl[{{4\pi^2\over3h}}\>\epsilon\Biggr]}^{1\over2}.
\end{equation}
For an energy budget of $1\:{\rm Js}^{-1}$, this gives a maximum
information rate of $2.03\times10^{17}$ bits per second.

A related issue is the calculation of the most information-efficient
broadcast transmission, where a broadcast in this context means a
transmission for which we do not know the location or size of the receiver
and therefore cannot guarantee that it will receive all, or even most, of
the photons transmitted.  In this case we again find that the most
efficient transmission is one which has the same spectrum as a
one-dimensional black body.

To summarize, we have shown that optimizing the information-efficiency
of a transmission which employs electromagnetic radiation as the
information carrier, with a fixed energy budget per unit time, gives
rise to a spectrum of intensities identical to that of black-body
radiation.  In fact, to an observer who is not familiar with the
encoding scheme used, an optimally efficient message of this type will
be entirely indistinguishable from naturally occurring black-body
radiation.  One might be tempted to say that perhaps one could tell
the two apart by spotting patterns within particular frequency bands
of the message, or by performing some other decomposition of the
signal.  This however is ruled out by the maximization of the Shannon
information; any regularities which would allow one to draw such a
distinction are necessarily the result of less than optimal encoding.
For the case where the transverse momentum of photons is not used to
encode information, or for broadcast transmissions, the spectrum is
indistinguishable from that of a one-dimensional black body.  We have
also shown that the characteristic temperature of the message is
simply related to the energy used to send it, and we have derived an
upper limit on the rate at which information can be transmitted in
both cases.

We end with some discussion and speculations.  First, we should point out
that the ideas outlined here constitute only a thought experiment.  By a
sequence of deductions we have placed an upper limit on the information
efficiency of electromagnetic transmissions, but we have not shown how to
achieve that upper limit.  Second, we speculate that results similar to
those presented here may apply to transmissions using other radiative media
as well.  Since many natural processes maximize the Gibbs--Boltzmann
entropy, they should give rise to spectra indistinguishable from optimally
efficient transmissions.  For instance, if we had a transmitter that can
emit any type of particle (rather than just photons), it seems plausible
that the optimal spectrum of particle types and energies would be that of
Hawking black-hole radiation~\cite{Hawking75}.

\def\refer#1#2#3#4#5#6{{\frenchspacing\sc#1}\hspace{4pt} #2\hspace{8pt}#3
  {\it#4} {\bf#5}, #6.}
\def\bookref#1#2#3#4{{\frenchspacing\sc#1}\hspace{4pt}
  #2\hspace{8pt}{\frenchspacing\it#3} #4.}


\begin{references}
%
\bibitem{Shannon49}
\bookref{Shannon, C. E. and Weaver, W.}{1949}{The Mathematical Theory of
  Communication.}{University of Illinois Press, Urbana}
%
\bibitem{CT91}
\bookref{Cover, T. M. and Thomas, J. A.}{1991}{Elements of Information
  Theory.}{John Wiley, New York}
%
\bibitem{Jaynes83}
\bookref{Jaynes, E. T.}{1983}{Papers on Probability, Statistics and
  Statistical Physics.}{Reidel, Dordrecht}
%
\bibitem{Grandy87}
\bookref{Grandy, W. T.}{1987}{Foundations of Statistical
    Mechanics.}{Reidel, Dordrecht}
%
\bibitem{Hawking75}
\refer{Hawking, S. W.}{1975}{Particle creation by black
holes.}{Commun. Math. Phys.}{43}{199--220}

\end{references}
\end{document}